\documentclass{article} 
\usepackage{arxiv}
\usepackage{amsmath} 
\usepackage{amsthm} 
\usepackage{graphicx} 
\usepackage{algorithmic} 
\usepackage{subfig} 
\usepackage{color}
\bibpunct{[}{]}{,}{n}{,}{;}

\def\N{\mbox{I\hspace{-.15em}N}}
\title{Rock-Paper-Scissors Random Walks on Temporal Multilayer Networks}
\author{Martin Gueuning \\
naXys, Universit\'e de Namur, Belgium\\
ICTEAM, Universit\'e catholique de Louvain, Belgium\\
\texttt{martin.gueuning@unamur.be}
\And
Sibo Cheng\\
\'{E}lectricit\'e de France R$\&$D\\
LIMSI, CNRS, Univ. Paris-Sud, Universit\'e Paris-Saclay, France\\
ICTEAM, Universit\'e catholique de Louvain, Belgium\\
\And
Renaud Lambiotte\\
Mathematical Institute, University of Oxford, UK~~~~~~~~~~~~~~~\\
\And
Jean-Charles Delvenne\\
ICTEAM, Universit\'e catholique de Louvain, Belgium\\
}
\begin{document}
\maketitle
\begin{abstract}
{We study diffusion on a multilayer network where the contact dynamics between the nodes is governed by
a random process and where the waiting time distribution differs for edges from different layers.
 We study the impact on a random walk of the competition that naturally emerges between the edges of the different layers. In opposition to previous studies which have imposed a priori inter-layer competition, the competition is here induced by the heterogeneity of the activity on the  different layers. 
 We first study the precedence relation between different edges and by extension between different layers, and show that it determines biased paths for the walker. We also discuss the emergence of cyclic, rock-paper-scissors random walks, when the precedence between layers is non-transitive. Finally, we numerically show the slowing-down effect due to the competition on a heterogeneous multilayer as the walker is likely to be trapped for a longer time either on a single layer, or on an oriented cycle .\\
 
}
\end{abstract}

\section{Introduction}

The study of random walks has a long tradition in network science \cite{Masuda2017}. Random walks are at the heart of many algorithms to uncover central nodes or communities of densely connected nodes, and they often serve as a first model to understand how the topology of a network affects diffusive processes. 
Random Walks have also been studied mathematically and numerically when the underlying topology is a network enriched with additional features. An important family of models consists of temporal networks  \cite{Holme2012,Guide2016}, where edges are dynamical entities and diffusion can only propagate when they are active. In this case, random walks are affected by the interplay between the network topology and the statistical properties of the edge dynamics \cite{Starnini2012,Hoffmann2013,Delvenne2015}. Another important family is made of multiplex networks \cite{Domenico2016,Kivela2014,Aleta2018}, where different types of connections exist between the nodes, as in social  networks \cite{szell2010multirelational,Magnani2013} or transportation networks \cite{Cardillo2013,Gallotti2015} for instance. Multiplex networks have a layered organisation and are usually represented as tensors or by means of a so-called supra-adjacency matrix.  Random walks have also been studied in this context, to uncover how the presence of multiple layers affects diffusion \cite{Domenico2013,Domenico2014} or to define generalized versions of Pagerank \cite{de2015ranking}.

The main objective of this work is to explore phenomena emerging in the case of  networks that are temporal and multiplex \cite{Starnini2017}. To do so, we extend a generic model of continuous-time random walk on  networks \cite{Hoffmann2013}, where the waiting time distribution for an edge, that is the time before an edge gets active, depends solely on its layer. We study the properties of the resulting stochastic process and show that the presence of temporal heterogeneities across layers results in a competition between them and biases in the random walk dynamics. Here competition means that edges in one layer may have a higher probability to be selected by a walker than edges in another layer, due to the statistical properties of their temporal ordering. In other words, competition between layers emerges due to the temporality of the graph, and not as a model parameter as in previous works \cite{Ding2017,Gomez-Gardenes2015,Kleineberg2016,Jang2015}. In addition, we show and explain some apparently counter-intuitive situations,  such as the emergence of a cyclic, rock-paper-scissors precedence between the layers. Note that the notion of non-transitivity is well-known in statistics, and that it has mostly focused on systems having a finite set of possible states, such as in non-transitive dice \cite{Gardner1970}. Our work can be seen as an extension to continuous variables and a study of its impact on diffusion over multilayer networks.
As a second step, we  explore numerically the impact of the above mechanism on the dynamics of a walker and, specifically, we study the coverage of a walker on temporal multilayer network.

\section{Random walks on multiplex temporal networks}
We consider a diffusion process defined as follows \cite{Masuda2017}. The system is made of a given network of nodes and edges, and a random walker jumping between neighbouring nodes according to a given protocol as we now discuss. In a standard framework of temporal networks, which is a natural model for contacts in social networks, we assume that when the random walker arrives on a node, it triggers on each incident outgoing edge an associated random waiting time for so-called `activation' of the edge, and the first edge to reach activation is then followed by the random walker. The process is restarted on the new node reached by the random walker. Here activation is therefore seen as an event of infinitesimal duration, allowing the passage of the random walker. The time at which each edge reaches activation is independently drawn from its own waiting time distribution.
The walk is called 
 active since waiting times are drawn once the walker arrives on a node, as opposite to the passive walk where contacts are taking place on edges regardless of the presence of a random walker. Note that, when the network is directed and has no short cycles,  a passive walk may be approximated by an active one provided that waiting time distributions are adapted accordingly \cite{Gueuning2017,Masuda2017}. 
Because only the first edge to reach activation is taken by the walker, there is a underlying competition between different edges. 
 Even though a general master equation can be derived  \cite{Hoffmann2013}, specific results have often been studied with homogeneous waiting time distribution \cite{Karsai2017,lala,Gueuning2015}.
 Here, we consider a 
  situation where edges in the same layer of the multiplex network have the same waiting time distribution but distributions may differ for edges in different layers. The system thus exhibits intra-layer homogeneity and inter-layer heterogeneity. 
\subsection{Emergence of biased paths}
Let us consider the trajectory of a random walker. Arriving on node $u$, the walker will leave through the first edge that reaches activation among the $k$ edges connected to $u$. 
Denote $T_{1},\ldots,T_{k}$ the random variables associated to the waiting times associated to each edge, with distributions $f_1(t),\ldots,f_k(t)$. The transition time $T$ of the walker, defined as the time before its next jump is given by:
\begin{eqnarray}
    T&=&\min{\left(T_{1},T_{2},....,T_{k}\right)} \label{eq:T}.
\end{eqnarray}

In our case, each edge belongs to a layer and it is thus natural to determine which layer will be selected by the walker. In particular, we will be interested in a notion of precedence between the random variables of the layers.
The random variable $A$ is said to precede the random variable $B$, written $A \prec B$, if $P(A<B)>0.5$, that is if the edge associated to $A$ is more likely to reach activation before the edge associated to $B$.\\

The existence of precedence relations translates into biased paths for the random walker traveling on the network since at each step of the walk some edges may be statistically more likely to be selected by the walker. Therefore, understanding the precedence relation between the random variables associated to the dynamics of the network is of paramount interest in order to understand the resulting diffusion. In the following, we will exemplify the somewhat counter-intuitive properties of precedence, 
 then we will numerically investigate its impact on the coverage of a random walker on a multilayer network with inter-layer heterogeneity.
 
\subsection{Basic properties of precedence }
\subsubsection{Rock-Paper-Scissors  }

\begin{figure}[!h]
\centering\includegraphics[scale=1]{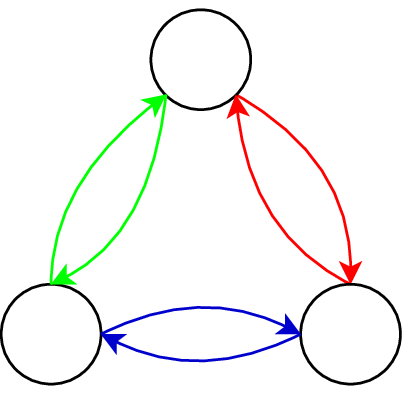} 
	\centering\includegraphics[scale=0.5]{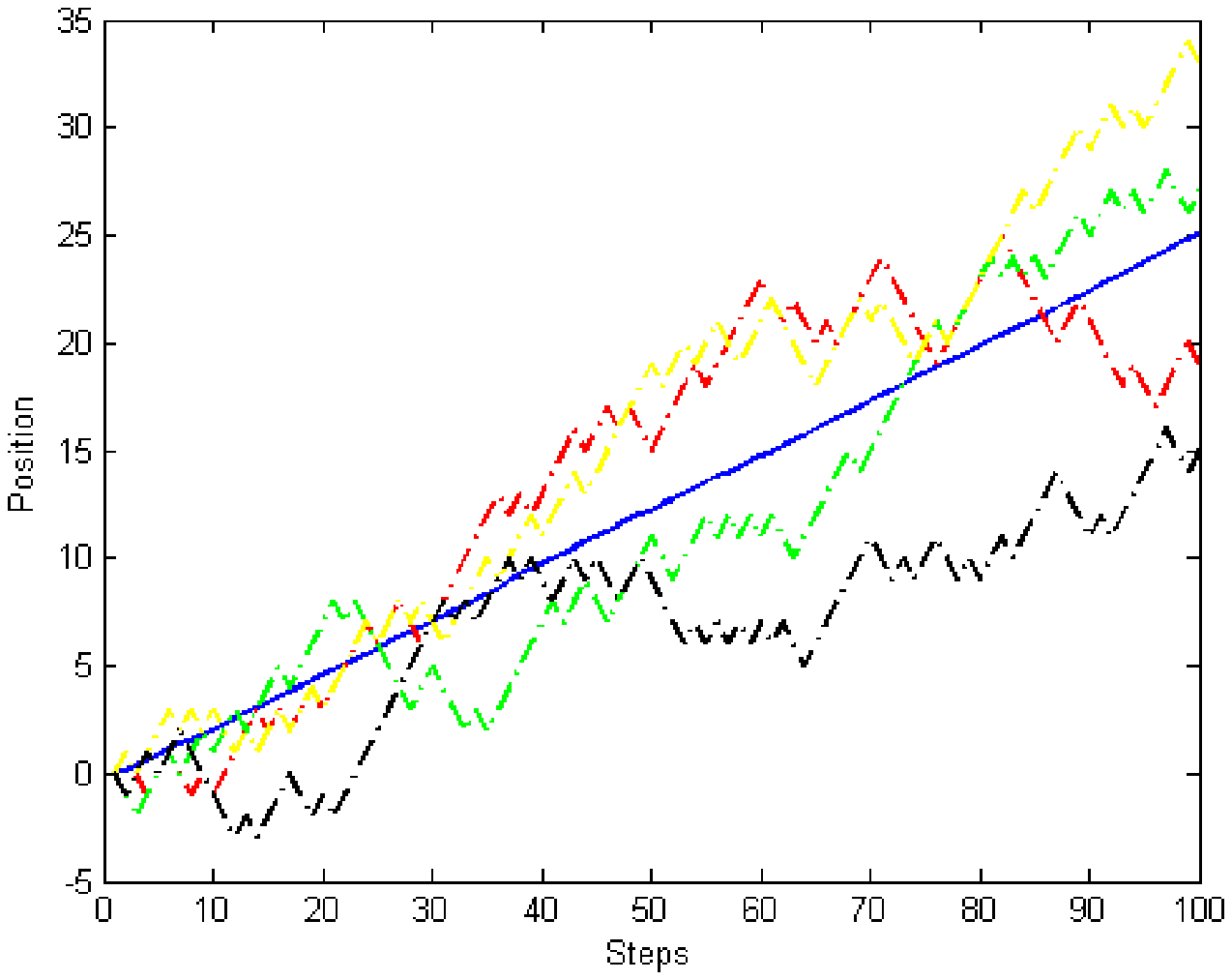} 
\caption{Numerical simulations of a random walk on a multilayer triangle. Each (colored) pairwise relation belongs to a different layer and distributions activity of the layers are given by the distribution of (\ref{distr}). The walker position is incremented $+1$ at each clockwise jump, and $-1$ at each anti-clockwise jump. On the right hand side, the four dash-dot lines represent four realisations of random walk starting at $0$ while the blue line represent the average over 1000 independent walks.  The precedence relation is not transitive as a random walker jumping on the triangle will have a tendency to jump clockwise.}
\label{fig:sim}
\end{figure}

One may see precedence as a relation of dominance between edges in competition to attract the random walker. As such one may expect transitivity. However it is not the case, and we encounter circular, rock-paper-scissors situations as follows.
We focus on the triangle network of Figure \ref{fig:sim} and the three following distributions with expectation equal to $1$:\\
\begin{eqnarray}\label{distr}
          X &=&\left\{
                \begin{array}{ll}
              U\left[0.65,0.75\right] &\quad \textrm{with probability} \quad 1-\frac{1}{\varphi}\\
                             U\left[0.3\varphi+0.65,0.3\varphi+0.75\right] &\quad \textrm{with probability} \quad \frac{1}{\varphi};
                 
                \end{array}
              \right.\notag\\ 
              &&\notag\\
                         Y&\sim&\quad ~U\left[0.9,1.1\right] ;\\ 
                         &&\notag\\
         Z&\sim&\left\{
                     \begin{array}{ll}
                          U~[0.75,0.85] &\quad\quad\quad \textrm{with probability} \quad \frac{1}{\varphi} \\
                          U\left[\frac{\varphi}{5}+0.95,\frac{\varphi}{5}+1.05\right] &\quad\quad\quad \textrm{with probability} \quad  1-\frac{1}{\varphi} ,
                         \end{array}
                       \right. \notag
\end{eqnarray}
where $U[a,b]$ stands for the uniform distribution on the interval $[a,b]$, and $\varphi=\frac{1+\sqrt{5}}{2}\approx 1.618$.

It is straightforward to show that $Y\prec X$, $Z\prec Y$ and $X\prec Z$. As a consequence, a walker jumping on the network will have a tendency to jump clockwise on the triangle, as illustrated numerically in Figure \ref{fig:sim}. 
The emergence of a circular flow leads to correlations between the successive edges on the random walk trajectory, even if edges are chosen independently at random at each step.
Competition becomes  more complex when more than two edges interact together. 
For instance, in the above example, even if the pairwise precedence is uniform, that is 
 \begin{eqnarray}
 P(Y<X)=P(Z<Y)=P(X<Z)=\frac{1}{\varphi},
 \end{eqnarray} 
 when considering the competition between 
3 different edges, one finds as $P(X<\min(Y,Z))>\frac{1}{3}$, and thus $X$ will tend to be favoured by the walker against the others two in the presence of the three types of edges simultaneously.

\newpage
\subsubsection{There is no most preceding edge }

It is clear that comparing either means or variances of two random variables is not sufficient to determine which one precedes the other, as the random variables $X$, $Y$ and $Z$ in the previous example share equal mean but have different variance.  
Even more, it is impossible to find a random variable that precedes any random variable with equal mean.
Without loss of generality, we  prove this statement for random variables with a mean of $1$.
First, we observe that for any random variable $X$ following distribution $f(t)$, it is possible to find a random variable $Y_n$ of distribution $g_n(t)$  such that $Y_n$ precedes $X$, by setting \begin{eqnarray}
g_n(t)= &  &\left\{
 \begin{array}{ll}
  \displaystyle\frac{n}{2} (n-1) & \mbox{if } t\in\left[\frac{1}{n}-\frac{1}{n^2}~,~\frac{1}{n}+\frac{1}{n^2}\right] \\
  \displaystyle\frac{1}{2 \sqrt{n}}&\mbox{if } t\in\left[n-1+\frac{1}{n}-\frac{1}{\sqrt{n}}~,~ n-1+\frac{1}{n}+\frac{1}{\sqrt{n}}\right], \\
 0 & \mbox{otherwise}
  \end{array}
 \right.
\end{eqnarray}
 where $n\in\N_0$ is chosen large enough. One possible choice of $n$ is such that $\frac{1}{n}+\frac{1}{n^2}<\beta$, where $\beta$ is the $10\%$ quantile of $f$. 
From the sequence of random variables $\left(Y_n\right)_{n\in \N}$ with distribution respectively given by $\left(g_n\right)_{n\in \N}$, it is possible to extract a subsequence of random variables $\left(Z_n\right)_{n\in \N}$ whose respective distributions have non-overlapping supports. Such a sequence is increasingly precedent, that is, for any $n \in \N$ we have $Z_{n+1} \prec Z_n$.

As the series $\left(g_n\right)_{n\in \N}$ does not converge to a probability distribution, there exists no most preceding random variable.
In terms of diffusion on multiplex networks, this results implies that there is no optimal distribution ensuring that a given layer always captures a majority of the random walk flow, independently of the distributions in the other layers.
After the choice of the other layers have been made, a layer may always find a  distribution that will allow it to be the most precedent.

\subsubsection{Layer Precedence and Node Out-degree}

The notion of precedence between random variables naturally extends to precedence between layers when waiting time distributions inside a layer are homogeneous but vary across layers. In this case, the layer $L_1$ precedes the layer $L_2$ at node $u$ if the walker sitting on $u$ is more likely to perform her next jump through an edge of $L_1$, that is if $L_1$ is more likely to capture the flow passing through $u$. Moreover, in the presence of several layers, $L_1$ precedes $L_2$ and $L_3$ jointly if $L_1$ precedes the artificial layer $L_2 L_3$ obtained by the union of the layers $L_2$ and $L_3$.

Precedence of layers at a node depends on its out-degree $k_i$ on each layer $L_i$, as the layer of the edge selected by a walker is determined by comparing the smallest time $m_i$ to reach activation on each layer $L_i$, where $m_i$ is the minimum of $k_i$ random variables with identical distributions associated to the layer $L_i$. In the simplest case of two duplicate layers $L_1$ and $L_2$ with waiting time on edges of the type $X$ and $Y$ respectively, that is $L_1$ and $L_2$ share the same nodes and have the same edges, the out-going edges consist of $k$ edges of each type, and $L_1$ precedes $L_2$ if   \begin{eqnarray}
\displaystyle\min_{j=1,\ldots,k}X_j \prec \displaystyle\min_{j=1  \ldots,k}Y_j ,
\end{eqnarray}
where $X_j$ and $Y_j$ are duplicates of random variables following the same distribution than $X$ and $Y$ respectively.

The layer precedence may therefore vary between nodes depending on their out-degree, in particular for high and low out-degree nodes. However, it is worth to note that there always exists a threshold out-degree $\nu^*$ above which one layer will always precede the other one. Indeed, the distribution $f$ is said to have a larger minimal weight than the distribution $g$ if there exists $\epsilon > 0$ such that $P(X<\epsilon) > P(Y<\epsilon)$ and $\forall ~0< \sigma < \epsilon$, $P(X<\sigma) \geq P(Y<\sigma)$, where $X$ and $Y$ are two random variables following the distribution $f$ and $g$ respectively. Then, there always exists a threshold out-degree $\nu^*$ above which the minimum of at least $\nu^*$ realizations of the distribution with the larger minimal weight will precede the other one.\\

\begin{figure}[h!]
	\centering\includegraphics[scale=0.45]{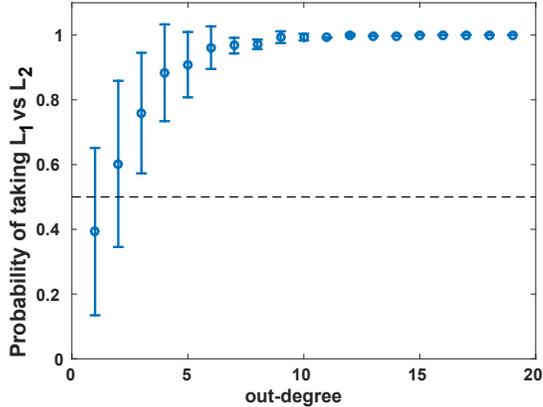}
	\caption{ We display the average probability ($\pm$ the standard deviation) of taking an edge of $L_1$ versus one of $L_2$, with respect to the out-degree of each node on the replicate layers. We observe a switch in the layers precedence: as the nodes of out-degree equal to one tend to favor the layer $L_2$, nodes of out-degree $\geq2$ favor the layer $L_1$. The distribution of the waiting times of the edges of $L_1$ and $L_2$ are the distributions of the random variables $X$ and $Y$ respectively, defined in Eq \ref{distr}. Dashed line corresponds to a probability equal to $0.5$. }
	\label{fig:switch_precedence}
\end{figure}

In order to investigate this effect in a real-world setting, we construct a multiplex network as follows.
We use a dataset of private messages sent on an online social network at the University of California Irvine \cite{Pietro2009}.
This typical social network is then duplicated to create a hypothetical two-layer social network, where each layer can be thought of as corresponding to a different medium of communication. 
Each layer is thus identical, and is composed of  $1899$ nodes and $20296$ edges. 
 The difference between the layers is induced by the choice of two different distributions of $X$ and $Y$ defined in Eq \ref{distr}, where $Y\prec X$. As $X$ has a larger minimal weight than $Y$, there exists a switch in the precedence between the variables of type $X$ and $Y$. In this case, it is straightforward to show that this switch occurs when at least two edges of each type are competing, that is \begin{eqnarray}\displaystyle\min_{i=1,\ldots,\nu}X_i \prec \displaystyle\min_{i=1  \ldots,\nu}Y_i~~ \forall \nu\leq2.\end{eqnarray} 
 Figure \ref{fig:switch_precedence} shows the probability of taking an edge of type $X$ instead of $Y$ with respect to the out-degree of a node in the static aggregated network (by construction, twice the out-degree in each layer), where each edge has two activation times associated to the random variable $X$ and $Y$ respectively. The numerical results confirm a switch of the precedence when the out-degree of the nodes increases.\\
\section{Impact of Inter-layer Heterogeneity on the Coverage of the Walker} \label{Mixing}

Finally, we  investigate numerically the impact of  competition in the case of diffusion in a multiplex social network with more than two layers. To do so, we use publicly available data introduced in \cite{Magnani2013} where the layers consist in five kinds of social  relationships between employees. In the numerical simulations, we  consider the three types of random variables defined in Eq.\ref{distr} and focus on the layers associated to Facebook, Leisure and Co-authorship relations. The structure of the graph thus corresponds to real-world interactions, while the inter-events time are chosen for the sake of illustration. Indeed, the presence of three layers and the choice of the specific distributions allow the emergence of properties of precedence we have previously shown, such as a rock-paper-scissors situation, and thus enables the investigation of its impact on the random walk.

\begin{figure}[h!]
	\centering
	\subfloat{
	\includegraphics[scale=0.35]{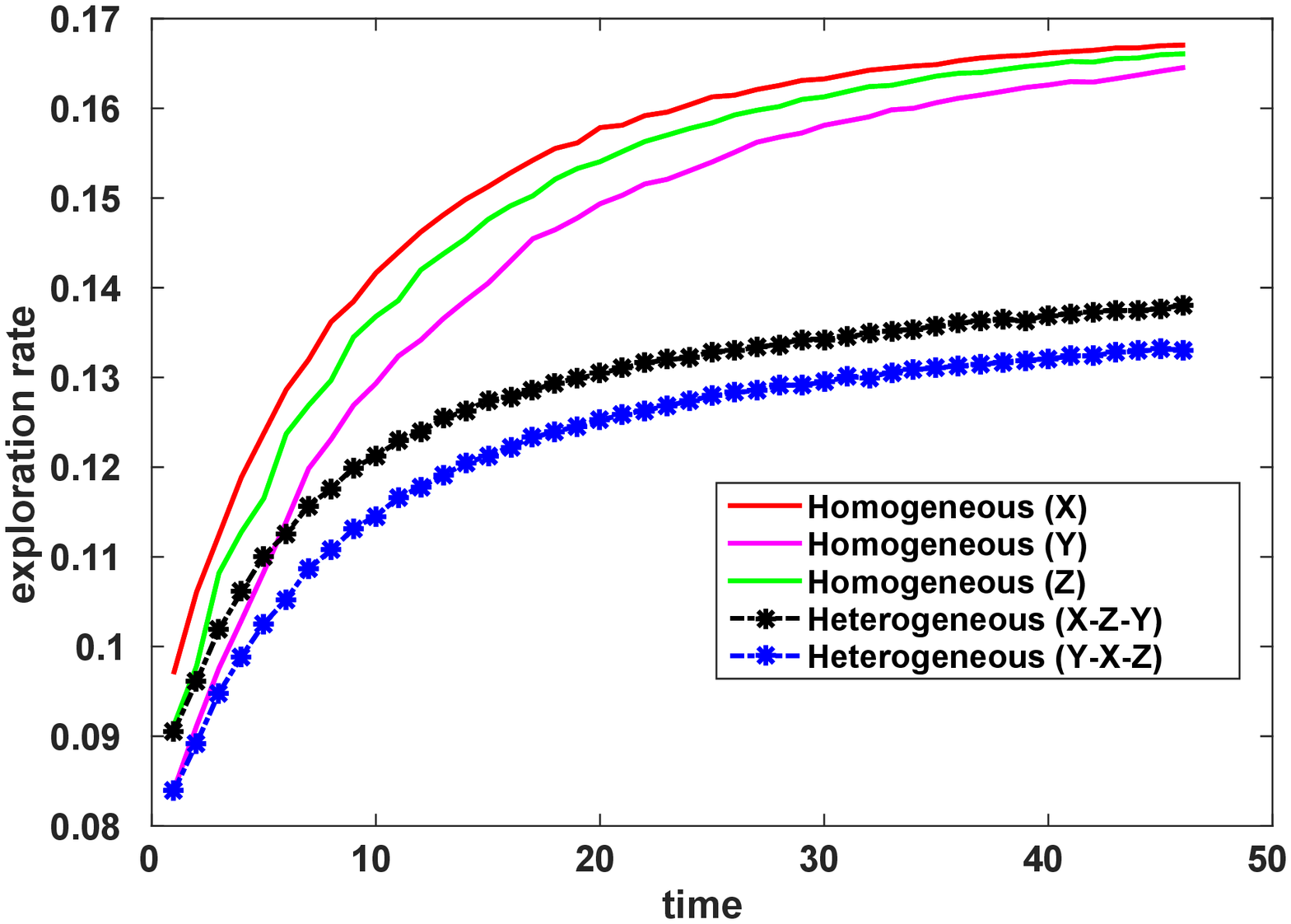}
	\label{fig:walk_real_crop}
	}
	\subfloat{
	\includegraphics[scale=0.34]{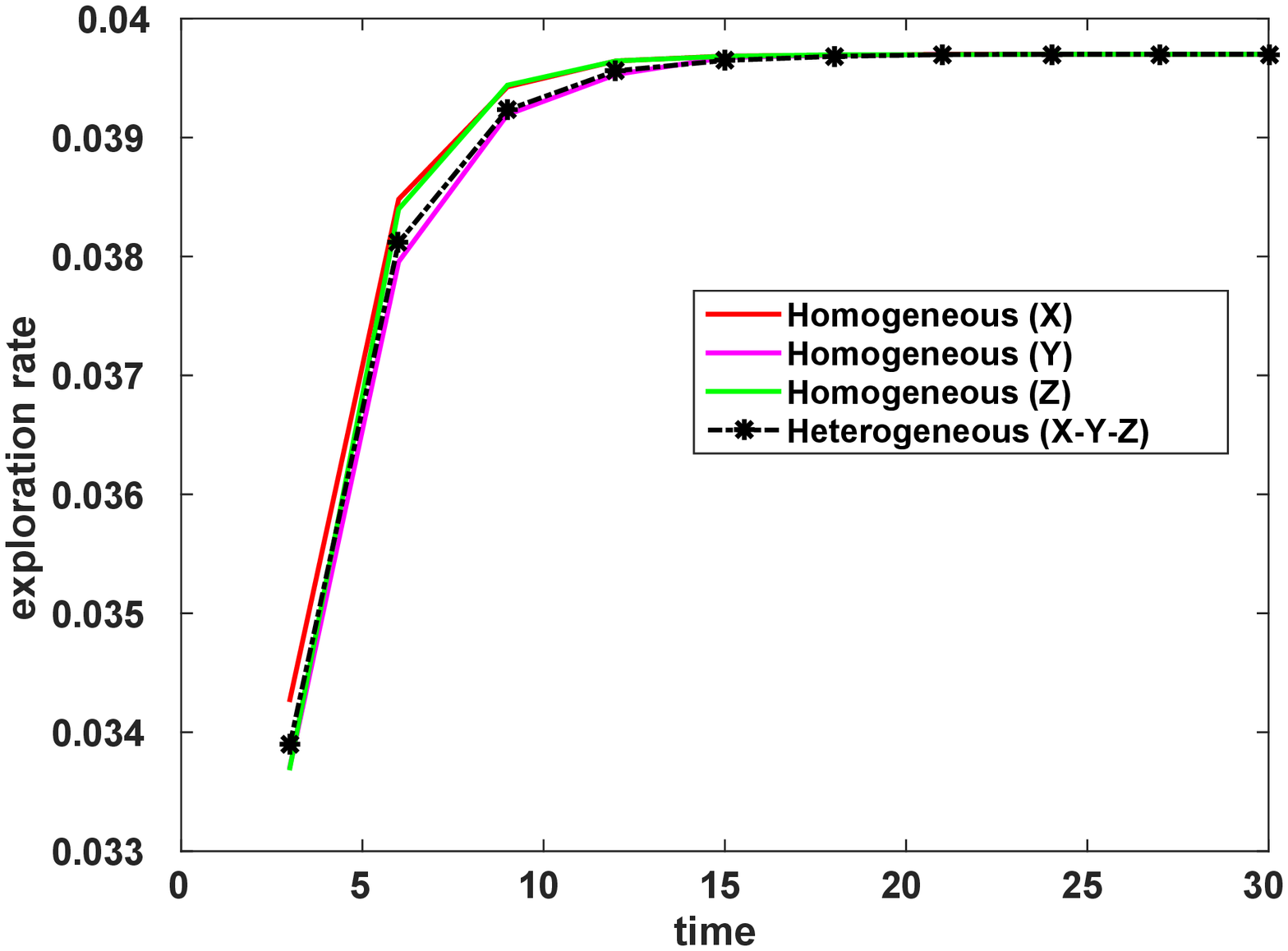}
	\label{fig:deg1} 
	} 
	\caption{Exploration rates over $100$ paths starting at each node of the multilayer network. Left: real-world network \cite{Magnani2013} with layers corresponding to Facebook, Leisure and Co-authorship relations respectively; Right: artificial network with out-degree of each node on each layer is set ton one. The waiting time distributions of the layers come from Eq.\ref{distr} and are provided in parenthesis. The exploration rate is larger under homogeneous inter-layer activity (plain lines) than under heterogeneous inter-layer activity (dashed lines) [left], except when out-degrees are low, as rock-paper-scissors situations emerge [right].\\
	} 
	\label{fig:exploration}
\end{figure}

We  compute numerically the exploration rate of a walker, or coverage, defined as the percentage of nodes visited by the walker  over time \cite{Domenico2014}. We always assume homogeneous waiting time distribution inside a given layer, and consider homogeneous as well as heterogeneous inter-layer distributions.

 A typical simulation is given in Figure \ref{fig:walk_real_crop}, where we observe that the exploration rate of a random walk tends to grow faster under inter-layer homogeneity compared to inter-layer heterogeneity, irrespective of the choice of the unique distribution. 
The slow-down induced by the inter-layer heterogeneity is mainly due to the fact that the flow is captured inside a single layer. Indeed, when graph density of the layers is large enough, one layer precedes the others two jointly, the one with edges of type $X$ in our example. Since the random walker has a tendency to stay in this layer, the walk will mainly take place on this layer, that is less connected than the aggregated network. However, when the graph density of the layers is weak, a rock-paper-scissors situation may arise, as the switch in precedence between layers might not occur for lower out-degree nodes. This emergence promotes the switch of the walker between different layers through lower out-degree nodes, resulting in a more efficient exploratory walk across the network, as illustrated in Figure \ref{fig:deg1} where each node has one outgoing edge in each layer, ensuring low density and homogeneous out-degree distribution. In this case, the exploration rate of the walk is similar to the exploration rate of an homogeneous inter-layer network (or monolayer network), which is the most efficient in terms of coverage rate.

 \section{Discussion}

The main purpose of this work is to investigate the competition between different layers of a multiplex network in situations when the network is temporal.   
 In our framework, edges activations are modeled as independent renewal processes, each layer being characterised by a different distribution, and  we highlight implications of the concept of precedence on diffusion.  In particular, we show that precedence may lead to biases between the different layers of the network.
 Despite the simplicity of the process, it may lead to counter-intuitive properties, such as non-transitivity,
  out-degree dependence, or rock-paper-scissors situation. Our numerical results also show that precedence may have important quantitative effects on the speed of diffusion on a multilayer network, as the precedence of one layer over others may hinder the number of edges available to the walker, and hence slow down its exploration of the graph.
We also show that high out-degree nodes are more prone to favor one single layer, while low out-degree nodes exhibit a different effect, and may lead to  a cyclic exploration between the layers. 
We study the impact of the precedence on an active random walk, however it is worth noticing that its impact on the passive random walk may lead to opposite bias towards layers. Indeed, in the presence of short cycles for the passive walker, an additional competition will emerge induced by the short cycles \cite{Petit2018}. For non-exponential distributions, this effect results in a backtracking bias towards or against the last traveled edges \cite{Gueuning2017}, typically leading to the emergence of short cycle patterns in human-related network \cite{Saramaki2015}.

  These results remain mostly mathematical as we used toy-model distributions instead of ones modelled on real-life data, and an important next step would be to test the resulting ideas in empirical data of multiplex networks where different layers are associated to different time scales, for instance between physical, mobile phone and social media interactions \cite{sekara2016fundamental}.

\section*{Acknowledgment}
This work has been partially supported by the Concerted  Research Action (ARC) supported by the Federation Wallonia-Brussels Contract ARC 14/19-060  ; and  Flagship European Research Area  Network  (FLAG-ERA)  Joint  Transnational  Call  “FuturICT  2.0”.
%
%

\bibliographystyle{imaiai}

\end{document}